\documentclass[review,final,sort&compress]{elsarticle}

\usepackage[mathscr]{eucal}
\usepackage{amsmath}
\usepackage{amssymb}
\usepackage{amsthm}
\usepackage{epsfig}
\usepackage{ifthen}
\usepackage{psfrag}
\usepackage[utf8]{inputenc}
\usepackage[english]{babel}
\addto\captionsenglish{}
\usepackage[T2A]{fontenc}
\usepackage{setspace}
\usepackage{url}
\usepackage[process=auto]{pstool}
\usepackage{hyperref}
\usepackage{stackrel}

\def\TODO#1{\marginpar{\setstretch{1}\vspace*{0mm}\hrule\strut\vphantom{p}\scriptsize #1\strut\vphantom{p}\hrule}}

\def\defe{\buildrel{\text{def}}\over=}
\def\tp{t^\prime}  
\def\prpr#1{#1^{\prime\prime}}
\def\thet#1{(\ref{#1})}
\def\veps{\varepsilon}
\def\Int{\int_{-\infty}^{+\infty}}
\def\IInt{\iint_{-\infty}^{+\infty}}
\def\VP{\operatorname {Vp}}
\def\PV{\operatorname {Vp}}
\def\const{\operatorname {const}}
\def\sign{\operatorname {sign}}
\def\Sqrt{\sqrt{\omega^2+k}}
\def\erf{\operatorname{erf}}
\def\S{{S}}
\def\C{{C}}
\def\defe{\buildrel{\text{def}}\over=}
\def\KK{{k_0}}
\def\Farg{\frac{\varkappa}{\sqrt{2\pi\mu}}}
\def\sign{\operatorname{sign}}
\def\const{\operatorname{const}}
\def\opm{{\omega^\pm}}
\def\Iint{\iint_{-\infty}^{+\infty}}  
\def\Np{\mathbf N^\prime}
\def\Rp{\mathbf R^\prime}
\def\bbU{\mathbf U}
\def\bbE{\mathbf E}
\def\bbF{\mathbf F}
\def\bbf{\mathbf f}
\def\bbN{\mathbf N}
\def\bbG{\mathbf G}
\def\bbL{\mathbf L}
\def\bbR{\mathbf R}
\def\bmR{\bmathcal R}
\def\bmF{\bmathcal F}
\def\mF{\mathcal F}
\def\bbr{\mathbf r}
\def\bbK{\mathbf K}
\def\bbi{\mathbf i}
\def\bbj{\mathbf j}
\def\ii{\bbi\,{\otimes}\,\bbi}
\def\dxi#1{{#1}_\xi}
\def\ddt#1{{#1}_{\tau\tau}}
\def\dt#1{{#1}_{\tau}}
\def\ddxit#1{{#1}_{\xi\tau}}
\def\ddxi#1{{#1}_{\xi\xi}}
\def\cD{\mathcal D}
\def\cK{\mathcal K}
\def\cEps{\mathcal E}
\def\pd#1#2{\dfrac{\partial#1}{\partial#2}}
\def\W{{W_0}}

\def\equ{equation\ }
\def\equs{equations\ }

\def\xi{x}
\def\tauu{t}

\def\Re{\operatorname{Re}}

\def\res{\operatorname{Res}}

\def\d{\mathrm d}

\def\i{\mathrm i}

\def\NEW#1{{#1}}

\journal{Journal of Sound and Vibration}

\begin{document}
\selectlanguage{english}
\begin{frontmatter}
\title{%
Non-stationary localized oscillations of an infinite
Bernoulli-Euler beam lying on the Winkler
foundation with a point elastic inhomogeneity of time-varying stiffness}
\author[ipme]{E.V.~Shishkina}
\ead{shishkina\_k@mail.ru}
\author[ipme,spbstu]{S.N.~Gavrilov\corref{mycorrespondingauthor}} 
\ead{serge@pdmi.ras.ru}
\cortext[mycorrespondingauthor]{Corresponding author}
\author[ipme]{Yu.A.~Mochalova}
\ead{yumochalova@yandex.ru}
\address[ipme]{Institute for Problems in Mechanical Engineering RAS, V.O., Bolshoy
pr. 61, St.~Petersburg, 199178, Russia}
\address[spbstu]{Peter the Great St.~Petersburg Polytechnic University,
Polytechnicheskaya str.~29, St.Petersburg, 195251, Russia}

\begin{abstract}
We consider non-stationary localized oscillations of an infinite
Bernoulli-Euler beam.
The beam lies on the Winkler foundation with a
point inhomogeneity (a concentrated spring with negative
time-varying stiffness).  In such a system with constant parameters (the
spring stiffness), under certain conditions a trapped mode of oscillation exists and is unique. Therefore, applying
a non-stationary external excitation to this system
can lead to the emergence of the beam oscillations localized near the inhomogeneity.
We provide an analytical description of non-stationary localized oscillations 
in the system with time-varying properties
using the asymptotic procedure based on successive application of two asymptotic
methods, namely the method of stationary phase and the method of multiple
scales. The obtained analytical results were verified by
independent numerical calculations.
The applicability of the analytical formulas was
demonstrated for various types of an external excitation and laws governing the
varying stiffness. In particular, we have shown that 
in the case when the trapped mode frequency approaches zero, localized low-frequency oscillations with increasing
amplitude precede the localized beam buckling.  The dependence of the amplitude of
such oscillations on its frequency is more complicated in comparison with the
case of a one degree of freedom system with time-varying stiffness.
\end{abstract}

\end{frontmatter}

\def\defe{\buildrel{\text{def}}\over=}
\def\tp{t^\prime}  
\def\prpr#1{#1^{\prime\prime}}
\def\thet#1{(\ref{#1})}
\def\veps{\varepsilon}
\def\Int{\int_{-\infty}^{+\infty}}
\def\IInt{\iint_{-\infty}^{+\infty}}
\def\VP{\operatorname {Vp}}
\def\PV{\operatorname {Vp}}
\def\const{\operatorname {const}}
\def\sign{\operatorname {sign}}
\def\Sqrt{\sqrt{\omega^2+k}}
\def\erf{\operatorname{erf}}
\def\S{{S}}
\def\C{{C}}
\def\defe{\buildrel{\text{def}}\over=}
\def\KK{{k_0}}
\def\Farg{\frac{\varkappa}{\sqrt{2\pi\mu}}}
\def\sign{\operatorname{sign}}
\def\const{\operatorname{const}}
\def\opm{{\omega^\pm}}
\def\Iint{\iint_{-\infty}^{+\infty}}  
\def\Np{\mathbf T^\prime}
\def\Rp{\mathbf R^\prime}
\def\bbU{\mathbf U}
\def\bbE{\mathbf E}
\def\bbF{\mathbf F}
\def\bbf{\mathbf f}
\def\bbN{\mathbf T}
\def\bbG{\mathbf G}
\def\bbL{\mathbf L}
\def\bbR{\mathbf R}
\def\bmR{\boldsymbol{\mathscr R}}
\def\bmF{\boldsymbol{\mathscr F}}
\def\mF{\mathscr F}
\def\bbr{\mathbf r}
\def\bbK{\mathbf K}
\def\bbi{\mathbf i}
\def\bbj{\mathbf j}
\def\ii{\bbi\,{\otimes}\,\bbi}
\def\dxi#1{{#1}_\xi}
\def\ddt#1{{#1}_{\tau\tau}}
\def\dt#1{{#1}_{\tau}}
\def\ddxit#1{{#1}_{\xi\tau}}
\def\ddxi#1{{#1}_{\xi\xi}}
\def\cD{\mathscr D}
\def\cK{\mathscr K}
\def\cEps{\mathscr E}
\def\pd#1#2{\dfrac{\partial#1}{\partial#2}}

\def\w{{u}}
\def\u{{w}}

\def\equ{Eq.~}
\def\equs{Eqs.~}

\theoremstyle{remark}
\newtheorem*{Remark}{\it Remark}
\newtheorem{remark}{\it Remark}

\def\TODO#1{\marginparwidth=15mm
\marginpar{\hrule\strut\vphantom{p}\scriptsize #1\strut\vphantom{p}\hrule}}

\def\varGamma{\mathscr D}

\def\GG{\mathscr G}

\section{Introduction}


In this paper we consider a mechanical system with mixed spectrum of natural
oscillations. Namely, we deal with an infinite Bernoulli-Euler beam lying
on the Winkler foundation with a
point inhomogeneity (a concentrated destabilizing spring with negative slowly time-varying
stiffness).  In the case of a constant spring stiffness the discrete part 
of the spectrum for such a system may contain
unique (positive) eigenvalue, which is less than the lowest
frequency for the beam on the uniform foundation. This
special natural frequency corresponds to a trapped mode of oscillation 
with eigenform localized near the spring. The phenomenon of
trapped modes was discovered in the theory of surface water waves
\cite{ursell1951trapping}. The examples of various mechanical systems, where
trapped modes can exist, can be found in studies
\cite{kaplunov1986torsional,
abramyan1992characteristics,
kaplunov1995simple,
abramyan1998trapping,
Gavrilov-2006-trans,
gavrilov2002etm,
alekseev2002vibration,
indeitsev2004localization,
kaplunov2008example,
motygin2008trapping,
nazarov2010sufficient,
pagneux2013trapped,
gavrilov2016trapped,
kaplunov2005localized,
Ind-book-R2E,
indeitsev2000resonance,
indeitsev2012motion,
wang2014vibration,
indeitsev2015localization,
indeitsev2016evolution,
gavrilov2017trapped}. 

\NEW{Note that similar problems for structures of finite length were considered 
in studies 
\cite{luongo2001mode,
abramyan2011oscillations,
abramian2014oscillations,
abramian2017oscillations}.  The problem concerning localized buckling 
of a finite structure is somewhat different since in the latter case we have
discrete spectrum of natural frequencies and countable set of modes.}

It is known \cite{kaplunov1986torsional,gavrilov2002etm,mciver2003excitation,
Ind-book-R2E,
indeitsev2016evolution,
gavrilov2017trapped,arXiv:1805.07382}
that applying non-stationary external excitation to a system
possessing trapped modes leads to the emergence of undamped oscillations
localized near the inhomogeneity. The large time asymptotics for such oscillation
can be found 
\cite{gavrilov2002etm,mciver2003excitation,
indeitsev2016evolution,
gavrilov2017trapped}
by means of the method of stationary phase \cite{fedoruk1977,nayfeh1993}.
\NEW{Gavrilov in \cite{gavrilov2002etm,Gavrilov-2006-trans} 
suggested an asymptotic procedure based on successive application of two asymptotic
methods, namely, the method of stationary phase \cite{fedoruk1977,nayfeh1993}
(which we use to describe the steady-state motion in a system with constant
properties)
and the method of multiple scales
\cite{nayfehperturbation,nayfeh1993} 
(which we use to describe a motion in a system with time-varying 
properties).
This approach allows us to investigate
non-stationary processes in perturbed systems, with slowly time-varying
parameters,
possessing trapped modes.} In 
studies 
\cite{gavrilov2002etm,Gavrilov-2006-trans}
the problem concerning non-uniform motion of a point mass along a taut string
on the Winkler foundation was considered and solved.
%
The asymptotic procedure suggested in 
\cite{gavrilov2002etm,Gavrilov-2006-trans} was successfully applied to
describe the evolution of the amplitude of the trapped mode of oscillations in a
taut string on the Winkler foundation with a point inertial inclusion with
time-varying mass \cite{indeitsev2016evolution} and in a taut string on the Winkler foundation
with a concentrated spring with negative time-varying stiffness
\cite{gavrilov2017trapped}. In preprint 
\cite{arXiv:1805.07382}
non-stationary localized oscillations of an infinite
string, with time-varying tension, lying on the Winkler
foundation with a point elastic inhomogeneity are considered. 
\NEW{The advantage of the approach is that it is not necessary to solve any
PDE describing a system with time-varying parameters. Applying the method of 
multiple scales results in a first order ODE with time-varying coefficients,
that can be solved analytically. The disadvantage is that we obtain the
solution at the point under inclusion
(or inhomogeneity)
only. The latter is not so important,
since the localized solution decays quickly with distance from the
inclusion. Another disadvantage is that we restrict ourselves
to the important particular case of the general
problem concerning non-stationary oscillation, where
any external excitation (and, in particular, non-zero
initial conditions) is applied only to the point under the inclusion.} 

All problems considered in previous papers 
\cite{gavrilov2002etm,Gavrilov-2006-trans,indeitsev2016evolution,gavrilov2017trapped,
arXiv:1805.07382} 
deal with a string on the Winkler foundation and 
are formulated for the Klein-Gordon-type PDE.
\NEW{In this paper we 
generalize the suggested approach to the case when the waveguide under
consideration is a Bernoulli-Euler beam. We choose 
the model of a spring with negative stiffness to obtain the simplest problem for a
beam possessing a trapped mode.
The aim is to prepare the corresponding mathematical apparatus for 
the problem for a compressed beam with time-varying compression 
and a distributed defect (a weakened zone) 
in the Winkler foundation 
(see \cite{indeitsev2015localization}, where the case of constant compression is
considered). The latter problem has a geophysical motivation \cite{gavrilov2016trapped}
and is much more sophisticated
in many aspects in comparison with the one considered in this
paper.}

The paper is organized as follows. In Section~\ref{Sec-formulation}, we
consider the formulation of the problem. In Section~\ref{Sec-unperturbed}, an
infinite Bernoulli-Euler beam lying on the Winkler foundation with a point
elastic inhomogeneity {\it of a constant stiffness} is considered.  To deal with such
a system we use the corresponding Green function in the frequency domain 
(see~\ref{app-G-f}).  In Section~\ref{Sec-unperturbed1} we consider the
system without an external excitation  and solve the spectral problem. We
demonstrate that the system under consideration can possess a mixed spectrum
of natural frequencies and find the necessary and sufficient conditions for
the existence of a trapped mode. In Section~\ref{Sec-unperturbed2} we consider the
system with an external excitation and use the method of stationary phase to
find non-vanishing localized non-stationary oscillations with the trapped mode
frequency. 
In Section~\ref{sec-perturbed} we consider the case of inhomogeneity (the concentrated
spring) with a slowly time-varying stiffness. We apply the asymptotic procedure
based on the method of multiple scales and obtain the asymptotic solution. To
find the unknown constants we use the results obtained in
Section~\ref{Sec-unperturbed2}.
In Section~\ref{Sec-numerics}, we compare the analytical results with
numerical ones obtained by means of solving  the Volterra integral equation
of second kind with its kernel expressed in terms of the corresponding Green
function in the time domain (see~\ref{app-G-t}). 
In the conclusion (Section~\ref{Sec-conclusion}), we discuss the basic
results of the paper.

\section{Mathematical formulation}
\label{Sec-formulation}
%
%

We consider transverse oscillation of an infinite Bernoulli-Euler beam on the Winkler
elastic foundation. The elastic foundation has a point inhomogeneity in the form of a concentrated spring of a negative stiffness.

The dimensionless
equations describing transverse oscillation of the beam can be written as
follows:
\begin{gather}
 \ddot{w} + w'''' + w  =  P(t)\delta(x), 
 \label{int-eq-base-eq}
 \\
 P(t)=-K(t)w(0,t)+p(t). 
 \label{eq-P(t)}
\end{gather}
Here 
$w(x,t)$ is the displacement of a point of the beam at the dimensionless
position $x$ and dimensionless time~$t$, 
$K(t)$ is the dimensionless stiffness of the concentrated
spring 
{(a given function of time)},
$P(t)$ is the unknown dimensionless force on the beam from
the spring, $p$ is the given dimensionless external force on the beam.
The initial conditions for Eq.~\eqref{int-eq-base-eq} can be
formulated in the following form, which is conventional for distributions (or
generalized functions) \cite{Vladimirov1971}:
\begin{equation}
w \big|_{\tauu<0}\equiv0.
\label{<0-SPRING}
\end{equation}

Integrating 
 \eqref{int-eq-base-eq}
over $\xi=0$ results in the following condition 
\begin{equation}
[w''']=-K(t)w(0,t)+p(t).
\label{1dof-SPRING-p}
\end{equation}
Here, and in what follows, $[\mu]\equiv\mu(\xi+0)-\mu(\xi-0)$ for any
arbitrary quantity~$\mu$.

\section{The unperturbed problem}
\label{Sec-unperturbed}
In this section we consider the case of concentrated spring of a constant
stiffness 
($K=\const$).
\subsection{Spectral problem}
\label{Sec-unperturbed1}
Put $p=0$ and consider the steady-state problem concerning the natural
oscillations of
the system described by Eqs.~\eqref{int-eq-base-eq}--\eqref{eq-P(t)}.
Take
\begin{equation}
w=\hat w(\xi)\,\exp(-\i\Omega \tauu).
\label{umega-SPRING}
\end{equation}
Let us show
that 
such a system possesses a mixed spectrum of natural frequencies.
There exists a continuous
spectrum of frequencies $|\Omega|\geq\Omega_\ast$, which lies higher than the cut-off (or boundary) frequency $\Omega_\ast=1$.
The modes corresponding to the frequencies from the continuous spectrum are
harmonic waves. Trapped modes correspond to the frequencies from the
discrete part of the spectrum, which lies lower than the cut-off frequency:
$0<|\Omega|<\Omega_\ast$. We want to demonstrate that for the problem under consideration
the only
one trapped mode can exist. 
Trapped modes are modes with finite energy, therefore, we require
$\hat w\to0,\ \hat w'\to0$ as $x\to\infty$, and
\begin{equation}
\Int\hat w^2\,\d\xi<\infty,\qquad
\Int\hat w'{}^2\,\d\xi<\infty.
\label{finite-energy}
\end{equation}
Substituting 
Eq.~\eqref{umega-SPRING}
into 
Eqs.~\eqref{int-eq-base-eq}--\eqref{eq-P(t)} 
and taking into account the expression for the Green function
in the frequency domain (see~\ref{app-G-f}, formulas 
\eqref{beam-spectrum-alt-r-gen}, 
\eqref{Gr_b}) results in 
\begin{equation}
\hat w(x)=-\frac {K\hat w(0)}2\, 
\frac{\mathrm{e}^{-a|x|}\cos{\big( a|x| -\pi/4 \big)}}
{(1-\Omega^2)^{3/4}}.
\label{pre-freq-eq}
\end{equation}
Calculating the right-hand side of
\eqref{pre-freq-eq}
at $x = 0$
yields 
the equation for eigen-frequencies of the discrete spectrum
\begin{equation}
K+{2\sqrt2}\,
{(1-\Omega^2)^{3/4}}
=0.
\label{fe}
\end{equation}
%
%
It is clear that frequency equation~\eqref{fe} can have real solutions only if
the stiffness of the inclusion is negative:  
\begin{equation}
K<0,
\label{K-negative}
\end{equation}
i.e.\ we deal with a destabilizing spring.
The solution of 
Eq.~\eqref{fe} is 
\begin{equation}
\Omega=\Omega_0\equiv\sqrt{1-\frac{|K|^{4/3}}4},
\label{c_omega}
\end{equation}
where $\Omega_0$ is the trapped mode frequency.
The corresponding localized form is proportional to the right-hand side of 
Eq.~\eqref{Gr_b}.

Thus, there exists the unique positive natural frequency less than the
cut-off frequency $\Omega_\ast = 1$ if and only if inequality 
\eqref{K-negative}
is true and
\begin{equation}
|K|<K_\ast\equiv 2\sqrt2.
\label{K-restrict}
\end{equation}
The critical value $K_\ast$
corresponds to the possibility of the localized buckling of 
the beam.

\subsection{Inhomogeneous non-stationary problem}
\label{Sec-unperturbed2}
Put now $p\neq0$. Applying to 
Eq.~\eqref{int-eq-base-eq}--\eqref{eq-P(t)}
the Fourier transform in time $\tauu$ results in 
\begin{equation}
 w_F'''' + (1-\Omega^2)w_F  =  (-Kw_F(0,\Omega)+p_F(\Omega)\big)\delta(x), 
 \label{ino-SPRING-non}
\end{equation}
where $w_F(x,\Omega),\ p_F(\Omega)$ are the Fourier transforms of
$w(\xi,\tauu)$ and $p(\tauu)$, respectively.
Using expressions 
\eqref{beam-spectrum-alt-r-gen},
\eqref{Gr_b} 
for the corresponding Green function
to resolve Eq.~\eqref{ino-SPRING-non} with respect to $w_F(0,\Omega)$ and
applying the inverse transform yields
\begin{equation}
w(0,\tauu)
=\frac1{2\pi}\Int
\frac{p_Fe^{-\i\Omega \tauu}\,\d\Omega}{
K+{2\sqrt2}\,
{(1-\Omega^2)^{3/4}}
}.
\label{before-stat-phase}
\end{equation}


Consider the case when $p(\tauu)$ is a vanishing as $\tauu\to\infty$ 
function such that its 
Fourier's transform 
$p_F(\Omega)$ does not have singular points 
on the real axis. 
Applying the residue theorem,  Jordan's lemma, and the method of stationary phase
to asymptotic
evaluation of the integral in
the right-hand side of 
\eqref{before-stat-phase} {results in}
\cite{fedoruk1977,Non-Stationary}
\begin{multline}
w(0,\tauu)=-\i\sum_{\bar{\Omega}=\pm\Omega_0-\i0}p_f(\bar{\Omega})
\res\left(
\frac{1}{
K+{2\sqrt2}\,
{(1-\Omega^2)^{3/4}}
},\bar{\Omega}\right)
\exp (-\i\bar\Omega t)+o(1),
\\
\tauu\to\infty.
\label{undamped-as-gengen}
\end{multline}
Here symbol $\res\big(f(\Omega),\bar\Omega\big)$ means the residue of function
$f(\Omega)$ at a pole $\Omega=\bar\Omega$.
The terms $-\i0$ in the expression for the poles
\begin{equation}
\bar\Omega=\pm\Omega_0-\i0
\label{poles}
\end{equation}
are taken in accordance with the principle of limit 	
absorption.
The asymptotic order of the error term in formula
\eqref{undamped-as-gengen} depends on the properties of $p_F$. One has 
\begin{equation}
\res\left(
\frac{1}{
K+{2\sqrt2}\,
{(1-\Omega^2)^{3/4}}
},\pm{\Omega_0}-\i0\right)
=\mp	
\frac{\sqrt[4]{1-\Omega_0^2}}{3{\sqrt 2}\,\Omega_0},
\end{equation}
thus
\begin{equation}
w(0,\tauu)=
\frac{{\sqrt 2}\sqrt[4]{1-\Omega_0^2}\,|p_F(\Omega_0)|}{3\Omega_0}
\sin\big(\Omega_0 \tauu-\arg p_F(\Omega_0)\big)+o(1), \quad
\tauu\to\infty.
\label{undamped-as-general}
\end{equation}
Taking into account 
\eqref{fe}, one can rewrite the last formula in the equivalent
form:
\begin{equation}
w(0,\tauu)=
\frac{2\sqrt[3]{|K|}\,|p_F(\Omega_0)|}{3\sqrt{4-{|K|^{4/3}} }}
\sin\big(\Omega_0 \tauu-\arg p_F(\Omega_0)\big)+o(1), \quad
\tauu\to\infty.
\label{undamped-as-general-equiv}
\end{equation}
Hence, for the large times, the
non-stationary response of the system under consideration is undamped
oscillations with the trapped mode frequency $\Omega_0$. 

\section{The perturbed problem}
\label{sec-perturbed}

Assume that the stiffness of the concentrated beam $K$ 
is slowly varying piecewise monotone
function of the dimensionless time $\tauu$:
$K=K(\epsilon \tauu)$.
Here $\epsilon$ is a formal small parameter.
We use an approach 
\cite{gavrilov2002etm,Gavrilov-2006-trans,indeitsev2016evolution}
based 
on the modification of the method of multiple scales
\cite{nayfehperturbation}
(Section~7.1.6) for equations with slowly varying coefficients.
The corresponding rigorous proofs, which validates such asymptotic approach in
the case of a one degree of freedom system, can be found in
\cite{feschenko1967eng}.
We look for the asymptotics for the solution under the following conditions:
\begin{itemize}
\item 
$
\epsilon=o(1), 
$
\item
$\tauu=O(\epsilon^{-1}),$
\item
$K(\epsilon \tauu)$ satisfies restriction 
\eqref{K-restrict}
for all $\tauu$.
\end{itemize}

To construct the particular solution of 
\eqref{int-eq-base-eq}, 
\eqref{eq-P(t)}
which describes the evolution of the trapped mode of oscillation 
in the case of slowly varying $K$, we  require that in the perturbed system
\begin{itemize}	
\item Frequency equation 
\eqref{fe}
for the trapped mode holds for all $\tauu$;
\item Dispersion relation 
\eqref{diss-rel-p=0}
at $\xi=\pm0$
holds for all $\tauu$.
\end{itemize}
Accordingly, we use the following ansatz ($\tauu>0$, $\xi\lessgtr 0$):
\begin{gather}
 w(\xi,\tauu)=W(X,T)\,\exp\varphi(\xi,\tauu),
 \label{slow-repr-SPRING}
 \\
 T=\epsilon \tauu,
 \quad
 X=\epsilon \xi,
 \\
 {\varphi}'=\i\omega(X,T),\quad
 \dot{\varphi}=-\i\Omega(X,T),
 \label{fast-phases-SPRING}\\
 W(X,T)=\sum_{j=0}^{\infty} \epsilon^{j}{W_j}(X,T).
 \label{W-series-SPRING}
\end{gather}
Here the amplitude $W(X,T)$, the wavenumber $\omega(X,T)$, and the frequency 
$\Omega(X,T)$ are the unknown functions to be defined in accordance with 
Eq.~\eqref{int-eq-base-eq}.
The variables $X,\ T,\ \varphi $ are assumed to be independent. 
We use the following representations for the
differential operators:
\begin{equation}
\begin{gathered}
 \dot{(\cdot)}=-
 \i\,\Omega \,\partial_{\varphi }+\epsilon\,\partial_{T},
 \\
 \ddot{(\cdot)}=-\Omega ^2\,\partial^2_{\varphi \varphi }
 -2\epsilon \i\,\Omega \,\partial^2_{\varphi T}
 -\epsilon \i\, {\Omega '}_T\,\partial_{\varphi }
 +O(\epsilon^2),
 \\
 \partial_{\xi}=
 \i\,\omega\,\partial_{\varphi}+\epsilon\,\partial_{X},
 \\
 \partial^2_{\xi\xi}=-\omega^2\,\partial^2_{\varphi\varphi}
 +\epsilon (2\i\,\omega\,\partial^2_{\varphi X}
 +\i\, {\omega}'_X\,\partial_{\varphi})
 +O(\epsilon^2),
 \\
 \partial^3_{\xi\xi\xi} = -\i\omega^3\partial^3_{\varphi\varphi\varphi}+
 \epsilon(-3 \omega ^{2} \partial^3_{\varphi\varphi X}   
 -3  \omega \omega'_X \partial^2_{\varphi\varphi})
 +O(\epsilon^2),
 \\
 \partial^4_{\xi\xi\xi\xi} = 
 \omega^4\partial^4_{\varphi\varphi\varphi\varphi}+
 \epsilon(-4\i \omega ^{3} \partial^4_{\varphi\varphi\varphi X}   
 -6\i  \omega    ^{2} \omega'_X \partial^3_{\varphi\varphi\varphi})
 +O(\epsilon^2).
\end{gathered}
\label{diff-operators}
\end{equation}
We require that $\omega(X,T)$ and $\Omega(X,T)$ satisfy dispersion relation
\eqref{diss-rel-p=0}
and equation 
\begin{equation}
{\Omega}'_X+{\omega}'_T=0
\label{dxx-SPRING}
\end{equation}
that follows from
\eqref{fast-phases-SPRING}.  Since in the case of a concentrated spring with constant
stiffness the
undamped oscillation can be described {by
Eq.~\eqref{undamped-as-general}},
we assume that
\begin{gather}
\Omega(\pm 0,T)=\Omega_0(T).
\label{Omega-Omega}
\end{gather}
Additionally, we require that 
\begin{gather}
[W]=0,\qquad 
[\varphi]=0.
\end{gather}
In Eq.~\eqref{Omega-Omega}
the right-hand side is defined in accordance with the frequency
equation \eqref{fe}, wherein $K=K(T)$.
The phase $\varphi(\xi,\tauu)$ should be defined by the formula 
\begin{equation}
 \varphi=\i\int(\omega\,\d\xi-\Omega\,\d\tauu).
\end{equation}

For large times, integrating formally Eq.~\eqref{int-eq-base-eq}
with respect to $\xi$ over the infinitesimal vicinity of 
$\xi=0$ taking into account  
\eqref{eq-P(t)},
one gets
\eqref{1dof-SPRING-p}, 
{wherein $p=0$}.
Now we substitute ansatz 
\eqref{slow-repr-SPRING}--\eqref{W-series-SPRING}
and representations \eqref{diff-operators}
into Eq.~\eqref{1dof-SPRING-p} and equate coefficients of like powers $\epsilon$. 
Taking into account frequency equation \eqref{fe},
and Eq.~\eqref{Omega-Omega},
one obtains that to the first approximation
\begin{equation}
[\omega^2 W_0{}'_X]=-[\omega\omega'_X]W_0
.
\label{1dof-1st-app-SPRING}
\end{equation}
Here
$
\omega=\omega_\pm
$ 
at $x=\pm0$, 
and $\omega_\pm$ is
defined by 
\eqref{a-a}.

On the other hand, the quantity in the left-hand side of 
\eqref{1dof-1st-app-SPRING} can be defined by consideration of 
Eq.~\eqref{int-eq-base-eq}
at $\xi=\pm0$. To do this,
we substitute ansatz 
\eqref{slow-repr-SPRING}--\eqref{W-series-SPRING}
and representations \eqref{diff-operators}
into Eq.~\eqref{int-eq-base-eq} 
and equate coefficients of like powers~$\epsilon$. 
Taking into account dispersion relation 
\eqref{diss-rel-p=0}
and Eq.~\eqref{Omega-Omega},
one obtains that to the first approximation
\begin{equation}
 4 \omega ^{3} W_0{}_X'  +
 6  \omega    ^{2} \omega'_X W_0
 +
 (2\Omega_0\,\W'_T+{\Omega_0}'_T\,\W)
 =0
\label{beam-1st-app-disp}
\end{equation}
at $x=\pm 0$. 
Equation
\eqref{beam-1st-app-disp} can be rewritten in the following form:
\begin{equation}
 \omega^2W_0{}'_X+
  \frac{\Omega_0}{2\omega}\,\W'_T
 + \left(\frac{1}{4\omega}\,{\Omega_0}'_T+
 \frac{3\omega\omega'_X}2  
 \right)\,\W
 =0.
\end{equation}
Calculating the jump results in
\begin{equation}
 [\omega^2W_0{}'_X]+
 \frac12\left[
 \frac{1}{\omega}
 \right]\,\Omega_0
 \W'_T
 + \left(
 \frac14
 \left[
 \frac{1}{\omega}
 \right]
 \,{\Omega_0}'_T+
 \frac{3[\omega\omega'_X]}2  
 \right)\,\W
 =0.
\end{equation}
Now, taking into account 
\eqref{1dof-1st-app-SPRING}, one gets 
\begin{equation}
\bar\W'_T+\left(\frac{{\Omega_0}'_T}{2\Omega_0}+\frac{[\omega\omega'_X]}{\Omega_0[1/\omega]}\right)\bar\W=0,
\label{eq-evol-p=0}
\end{equation}
where 
$\bar W_0(T) \equiv W_0(0,T)$.

Due to \eqref{dxx-SPRING}
one has
\begin{equation}
 {\omega}'_X={\omega}'_\Omega\,\Omega'_X=
 -{\omega}'_\Omega\,{\omega}'_T=-({\omega}'_\Omega)^2\Omega'_T,
\label{Omega-diff-X}
\end{equation}
where according to dispersion relation 
\begin{equation}
\omega'_{\Omega}=\frac{\Omega_0}{2\omega^3}.
\end{equation}
Hence,
\begin{equation}
\omega'_X=-\frac{\Omega_0^2}{4\omega^6}{\Omega_0}'_T,
\label{omega-diff-X-p=0}
\end{equation}
and 
\begin{equation}
\left[\omega\omega'_X
\right]=-\frac{1}{4}\Omega_0^2\Omega_0{}'_T\left[\frac{1}{\omega^5}\right].
\label{omega_omega-diff-X-p=0-pre}
\end{equation}
Taking into account the dispersion relation 
\eqref{diss-rel-p=0}, Eq.~\eqref{omega_omega-diff-X-p=0-pre}
can be rewritten in the following form:
\begin{equation}
\left[\omega\omega'_X
\right]=\frac{1}{4}\frac{\Omega_0^2{\Omega_0}'_T}{1-\Omega_0^2}\left[\frac{1}{\omega}\right].
\label{omega_omega-diff-X-p=0}
\end{equation}
Now we substitute the equation obtained into Eq.~\eqref{eq-evol-p=0}, and
finally obtain 
the first approximation equation
%
%
%
\begin{equation}
\bar\W'_T+\left(\frac{{\Omega_0}'_T}{2\Omega_0}+\frac{1}{4}\frac{\Omega_0{\Omega_0}'_T}{1-\Omega_0^2}\right)\bar\W=0.
\label{eq-evol-p=0-1}
\end{equation}
The general solution of the above equation is
\begin{equation}
\bar\W=C_0\exp\left(-\int{\frac{{\Omega_0}'_T}{2\Omega_0}}\,\d
T\right)\exp\left(-\int{\frac{1}{4}\frac{\Omega_0{\Omega_0}'_T}{1-\Omega_0^2}}\,\d
T\right),
\end{equation}
where $C_0$ is an arbitrary constant.
One has
\begin{equation}
\int{\frac{{\Omega_0}'_T}{2\Omega_0}}\,\d T=\frac{1}{2}\ln\Omega_0,
\end{equation}
and
\begin{equation}
\int{\frac{1}{4}\frac{\Omega_0{\Omega_0}'_T}{1-\Omega_0^2}}\,\d T=\frac{1}{8}\int
\frac{d\Omega_0^2}{1-\Omega_0^2}=-\frac{1}{8}\ln(1-\Omega_0^2).
\end{equation}
Thus, one obtains:
\begin{equation}
\bar\W=C_0\exp\left(-\frac{1}{2}\ln\Omega_0  \right)\exp \left(
\frac{1}{8}\ln(1-\Omega_0^2)
\right)=C_0\frac{\sqrt[8]{1-\Omega_0^2}}{\sqrt{\Omega_0}}.
\label{beam-analytic-evoK}
\end{equation}
Taking into account 
\eqref{fe}, one can rewrite the last formula in the equivalent
form:
\begin{equation}
\bar {W}_0
=C_0\frac{\sqrt[4]2\sqrt[6]{|K|}}{\sqrt[4]{4-{|K|^{4/3}}}}.
\label{sol-final-SPRING-var-T0-mod}
\end{equation}

One can see that $\Omega_0 \to 0$ if $K \to 2\sqrt{2}$.
When  $\Omega_0 \to 0$,
\begin{equation}
\bar {W}_0 = \frac{C_0}{\Omega_0^{1/2}}+o(1).
\label{A-propto-0}
\end{equation}
This result is analogous to the
classical result for a one degree of freedom system 
\begin{equation}
\ddot y+\hat\varOmega^2(\epsilon t) y =0, 
\label{1dof}
\end{equation}
where the following formula 
\begin{equation}
Y\propto \frac{1}{\hat\varOmega^{1/2}}
\end{equation}
for the amplitude of a natural oscillations $Y$ is valid 
(the Liouville -- Green approximation \cite{nayfehperturbation}). 
On the other hand, unlike one degree of freedom system \eqref{1dof}, for
the system under consideration, formula 
\eqref{A-propto-0}
is valid only in the limiting case $\Omega_0 \to +0$. 
For finite $\Omega_0$ the dependence for the amplitude is more complicated and 
is given by \eqref{beam-analytic-evoK}.
The analogous results were obtained for other similar problems (for a taut string on
the Winkler foundation with point inhomogeneity 
\cite{gavrilov2017trapped,arXiv:1805.07382}).

Combining the solution in the form of 
Eqs.~\eqref{slow-repr-SPRING}--\eqref{W-series-SPRING}
with its complex conjugate,
we get the non-stationary solution as the following ansatz:
\begin{equation}
w(0,\tauu)\sim
\bar W_0\big(\Omega_0(T)\big)\sin\left(\int_0^T\Omega_0(T)\,\d T-D_0\right),
\label{anzatz-final}
\end{equation}
where $\bar W_0$ is defined by 
\eqref{beam-analytic-evoK}.
The unknown constants $C_0$ and $D_0$ should be defined by equating the
right-hand sides of 
\eqref{undamped-as-general} and 
\eqref{anzatz-final} taken at $\tauu=0$. This yields
\begin{gather}	
C_0=
\frac{\sqrt2\sqrt[8]{1-\Omega_0^2(0)}}{3\sqrt{\Omega_0(0)}}
\,\big|p_F\big(\Omega_0(0)\big)\big|
=\frac{2^{3/4}\sqrt[6]{|K(0)|}}{3\sqrt[4]{4-{|K(0)|^{4/3}}}}\,\big|p_F\big(\Omega_0(0)\big)\big|,
\label{C0-an}
\\
D_0=\arg p_F\big(\Omega_0(0)\big).
\label{D0-an}
\end{gather}
In the particular case $p=\delta(\tauu)$ that corresponds to the initial
conditions in the classical form
\begin{equation}
w(0,0)=0,\qquad \dot w(0,0)=1
\end{equation}
one has
\begin{gather}	
C_0=\frac{\sqrt2\sqrt[8]{1-\Omega_0^2(0)}}{3\sqrt{\Omega_0(0)}},
\label{C0-an-simple}
\\
D_0=0.
\label{D0-an-simple}
\end{gather}
In the particular case  
\begin{equation}
p(\tauu)=H(\tauu)\exp(-\lambda \tauu),
\label{MASS:f-exp}
\end{equation}
where 
$H(\tauu)$ is the Heaviside function, and 
$\lambda=\mathrm{const}>0$, one has
\begin{gather}
p_F\big(\Omega_0(0)\big)=\frac{1}{
\lambda-\i\Omega_0(0)},
\label{psi-exp}
\\
\displaybreak[0]
\big|p_F\big(\Omega_0(0)\big)\big|=\frac{1}{\sqrt{\lambda^2+\Omega_0^2(0)}},
\label{psi-exp-abs}
\\
\displaybreak[0]
\arg\big(p_F\big(\Omega_0(0)\big)\big)=\arctan \frac{\Omega_0(0)}\lambda.
\label{psi-exp-arg}
\end{gather}%

\section{Numerics}
\label{Sec-numerics}
The solution satisfying Eq.~\eqref{int-eq-base-eq} and initial conditions 
\eqref{<0-SPRING} can be
written in the following form
\begin{equation}
w(x,t)=\int_{0}^{t} P(\tau)\mathscr G(t-\tau,x)\,\d\tau,
\end{equation}
where $\mathscr G$ is the fundamental solution 
(\ref{app-G-t}, formula \eqref{fs-td}).
On the other hand, from Eq.~\eqref{eq-P(t)} one obtains
\begin{equation}
w(0,t)=-K^{-1}(P(t)-p(t)).
\end{equation}
Thus, it is easy to get the following Volterra integral equation of the second
kind for $P(t)$:
\begin{equation}
P(t)=p(t)+K\int_{0}^{t} P(\tau)\mathscr G(t-\tau,0)\,\d\tau.
\label{volterra-beam}
\end{equation}
To solve integral equation 
\eqref{volterra-beam} we approximate the integral in the right-hand side
according to the trapezoidal rule \cite{book:811629}
and reduce the problem to a
system of linear equations with a triangular matrix. To perform the numerical
calculations we use 
{\sc SciPy} software.
All numerical results below are obtained for the following choice of the time step 
\begin{equation}
\Delta\tau=0.025. 
\label{Deltas}
\end{equation}
Calculating the numerical solutions, which corresponds to $p=\delta(\tauu)$, we
approximate the Dirac delta-function as follows: 
\begin{equation}
p=\tauu_0^{-1}(H(\tauu)-H(\tauu-\tauu_0)),
\end{equation}
where $H(\tauu)$ is the Heaviside function.

\begin{figure}[p]
\begin{center}
\includegraphics[width=\textwidth]{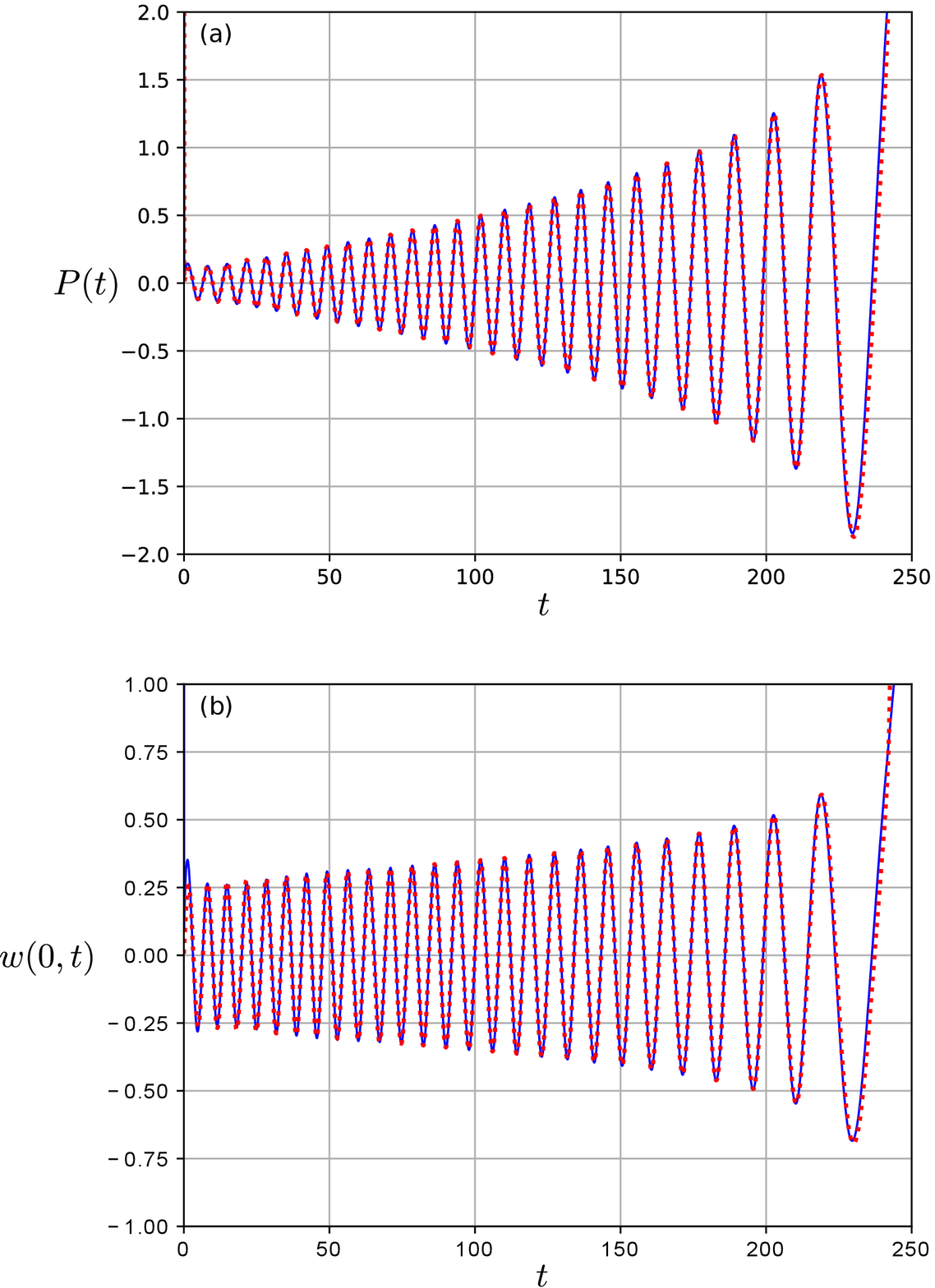}
\end{center}
\caption{
Comparing the analytical solution 
\eqref{eq-P(t)},
\eqref{sol-final-SPRING-var-T0-mod},
\eqref{anzatz-final},
\eqref{C0-an-simple},
\eqref{D0-an-simple}
obtained for $p=\delta(\tauu)$ 
(the red dotted line) and the numerical solution obtained for 
$p(\tauu)=\tauu_0^{-1}(H(\tauu)-H(\tauu-\tauu_0))$
(the blue solid line) in the case 
$K(\epsilon \tauu)={-0.4-\epsilon \tauu}$. Here $\epsilon=0.01,\ \tauu_0=0.1$.
The localized
buckling occurs at $\tauu\simeq242$.
\NEW{(a) the force $P(t)$, (b) the displacement $w(0,t)$}.
}
\label{test_k1.eps}
\end{figure}

\begin{figure}[p]
\begin{center}
\includegraphics[width=\textwidth]{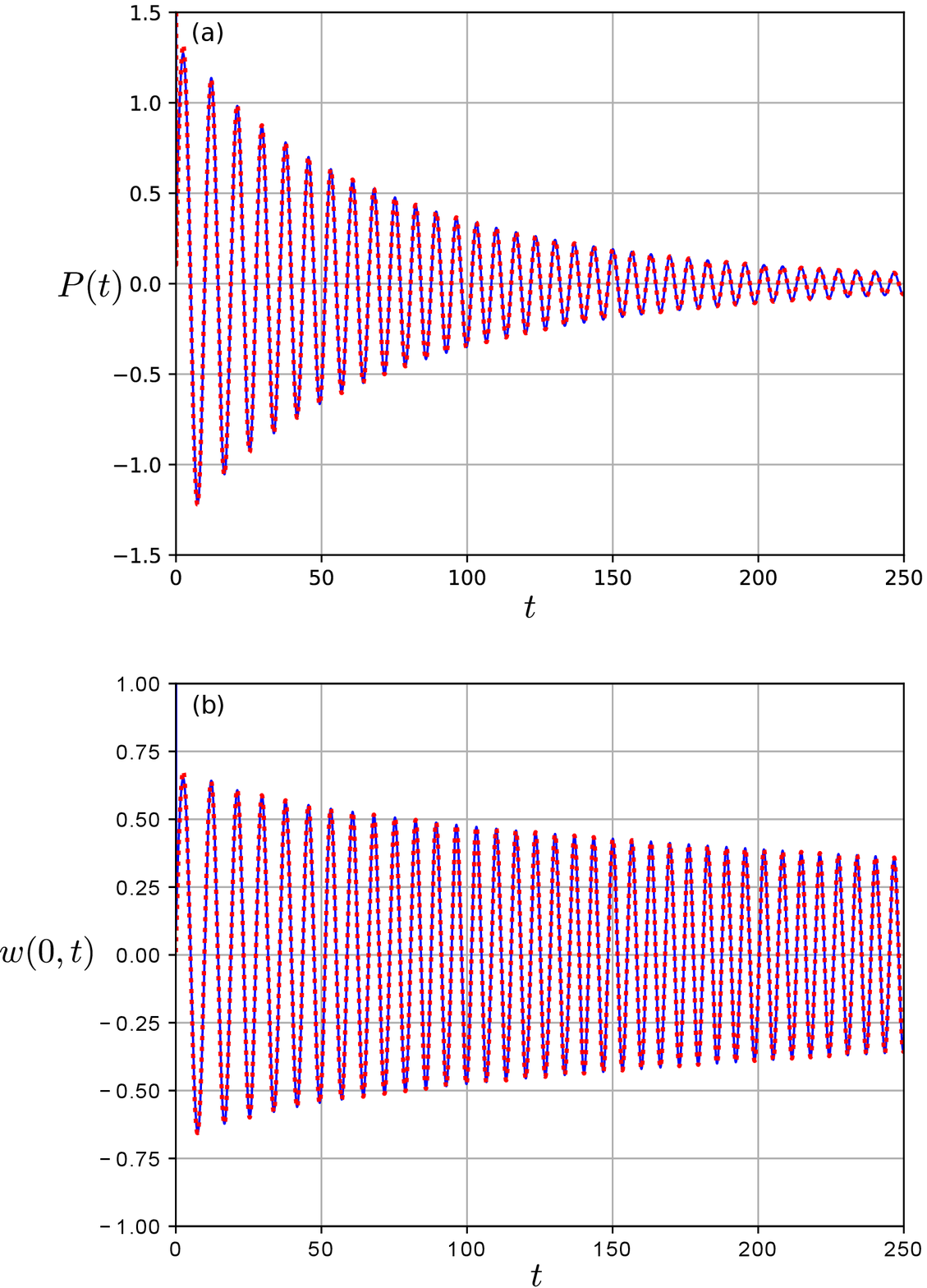}
\end{center}
\caption{
Comparing the analytical solution 
\eqref{eq-P(t)},
\eqref{sol-final-SPRING-var-T0-mod},
\eqref{anzatz-final},
\eqref{C0-an-simple},
\eqref{D0-an-simple}
obtained for $p=\delta(\tauu)$ 
(the red dotted line) and the numerical solution obtained for 
$p(\tauu)=\tauu_0^{-1}(H(\tauu)-H(\tauu-\tauu_0))$
(the blue solid line) in the case 
$K(\epsilon \tauu)={-2.0\exp(-\epsilon \tauu)}$. Here $\epsilon=0.01,\
\tauu_0=0.1$.
\NEW{(a) the force $P(t)$, (b) the displacement $w(0,t)$}.
}
\label{test_k2.eps}
\end{figure}

\begin{figure}[p]
\begin{center}
\includegraphics[width=\textwidth]{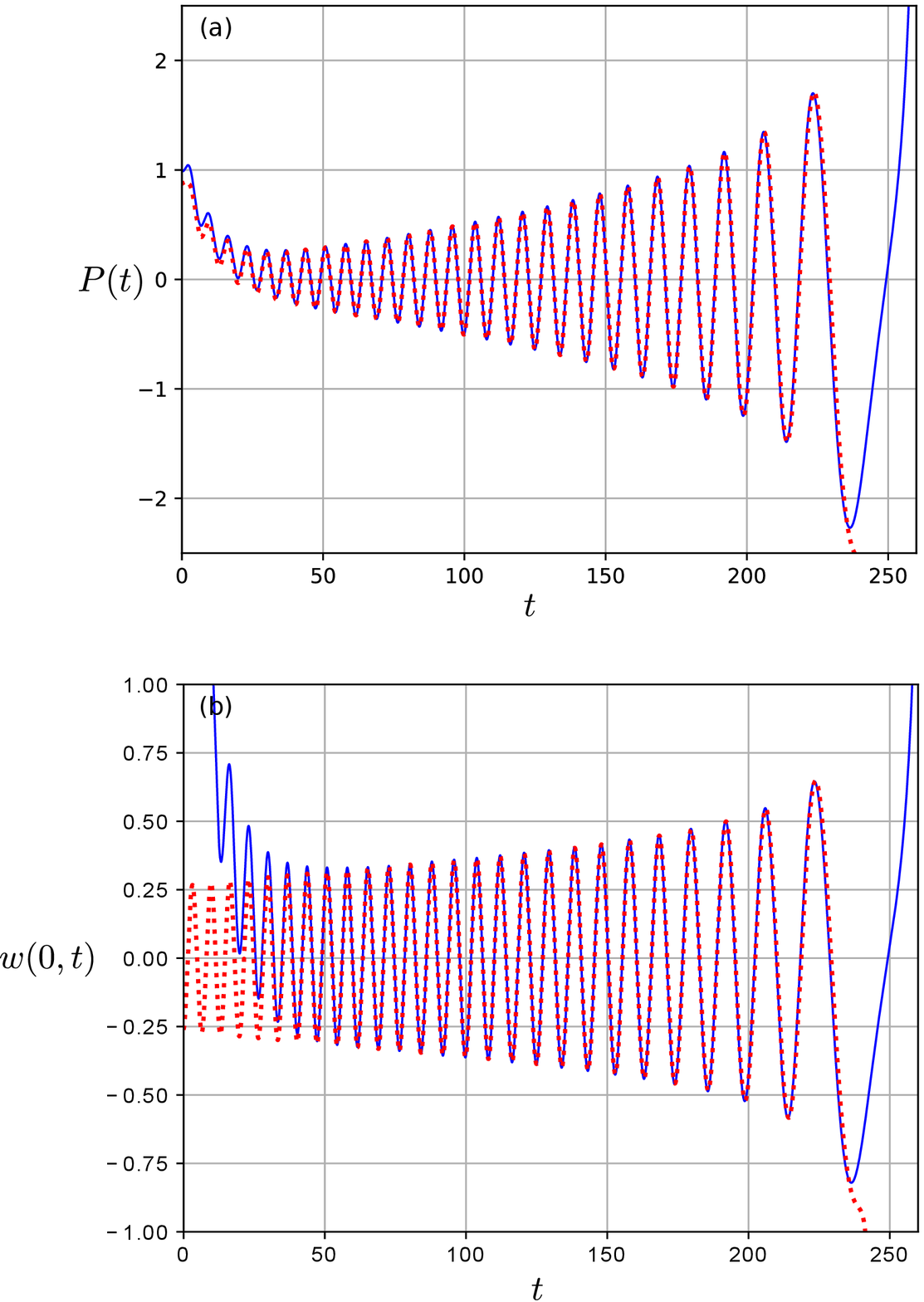}
\end{center}
\caption{
Comparing the analytical solution 
\eqref{sol-final-SPRING-var-T0-mod},
\eqref{anzatz-final},
\eqref{C0-an},
\eqref{D0-an},
\eqref{psi-exp-abs},
\eqref{psi-exp-arg}
(the red dashed line) and the numerical solution obtained for 
$p=H(\tauu)\exp(-\lambda\tauu)$ 
(the blue solid line) in the case 
$K(\epsilon \tauu)={-0.4-\epsilon \tauu}$.
Here $\epsilon=0.01,\ \lambda=0.1$.
The localized
buckling occurs at $\tauu\simeq242$.
\NEW{(a) the force $P(t)$, (b) the displacement $w(0,t)$}.
}
\label{test_k3.eps}
\end{figure}

A comparison between the analytical solution for $P(t)$ and $w(0,t)$
(given by formulas 
\eqref{eq-P(t)},
\eqref{sol-final-SPRING-var-T0-mod},
\eqref{anzatz-final},
\eqref{C0-an},
\eqref{D0-an}) and the numerical one
is presented
in Figures~\ref{test_k1.eps}--\ref{test_k3.eps}.
In Figure~\ref{test_k1.eps} we compare the results obtained for the case of
$p=\delta(\tauu)$ and monotonically increasing $|K(\epsilon \tauu)|$. The asymptotic
solution approaches the numeric one very quickly.
The localized
buckling occurs at $\tauu\simeq242$ that corresponds to the critical value
\eqref{K-restrict}.

In Figure~\ref{test_k2.eps} we compare the results obtained for the case of
$p=\delta(\tauu)$ and monotonically decreasing (vanishing) $|K(\epsilon \tauu)|$. 
Again the asymptotic
solution approaches the numeric one very quickly.

In Figure~\ref{test_k3.eps} we compare the results obtained for the case of
$p(\tauu)=H(\tauu)\exp(-\lambda\tauu)$ 
and monotonically increasing $|K(\epsilon \tauu)|$.
Since $\lambda=0.1$ is taken small enough, the method of the stationary phase 
gives a reasonable result only after some time ($\tauu\approx40$). After that
time the analytical solution approaches the numerical one. 

\section{Conclusion}
\label{Sec-conclusion}
In the paper we consider a non-stationary localized oscillations of an
infinite Bernoulli-Euler beam lying on the Winkler
foundation with a point inhomogeneity (a concentrated spring with negative slowly
time-varying stiffness.
In the case of the concentrated spring with a constant stiffness a 
trapped mode of oscillations exists and is unique if and only if conditions 
\eqref{K-negative}, \eqref{K-restrict} are satisfied. The existence of a trapped mode
leads to the possibility of the wave localization near the inhomogeneity 
in the Winkler foundation.
Applying a vanishing as $\tauu\to\infty$ external excitation to the
point of the beam under the inclusion leads to the emergence of undamped
free oscillations localized near the spring.
The most important result of the paper is analytical formulas 
\eqref{sol-final-SPRING-var-T0-mod},
\eqref{anzatz-final},
\eqref{C0-an},
\eqref{D0-an},
which allow us to describe for large times such non-stationary localized
oscillations in the case of slowly time-varying spring stiffness.
The obtained
analytical results were verified by independent numerical calculations
based on the Volterra integral equation of the second kind
\eqref{volterra-beam}, which is
equivalent to the initial value problem 
\eqref{int-eq-base-eq}--\eqref{<0-SPRING}.
The applicability of the analytical formulas was demonstrated for various types
of external excitation and laws governing the varying spring stiffness
(see Figures \ref{test_k1.eps}--\ref{test_k3.eps}).

We also have shown that localized low-frequency oscillations with 
increasing amplitude precede the localized beam buckling (see 
\eqref{A-propto-0}). However, unlike the case \eqref{1dof} of a one degree of freedom
system with time-varying stiffness,
in the framework of the problem under consideration
formula \eqref{A-propto-0} is correct only in
the limiting case, where the frequency of localized oscillations approaches zero. 
The analogous results were obtained for other similar problems (for a taut string on
the Winkler foundation with point inhomogeneity 
\cite{gavrilov2017trapped,arXiv:1805.07382}).
\NEW{In the paper we consider the pre-buckling stage only. This is a rather common
approach to investigate the dynamical stability.
The post-buckling is outside of the scope
of this paper.}

\NEW{Finally, it may be noted that the results of the paper can be used when
constructing dynamic materials
\cite{lurie2007introduction,rousseau2011elements},
and, in particular, dynamic
metamaterials
\cite{chronopoulos2015enhancement,pasternak2014chains,oyelade2017dynamics,
grekova2018dd}. 
In this case a dedicated controllable artificial device 
\cite{oyelade2017dynamics,huang2014vibration,ahn2014active,li2013negative,wu2014analysis}
can play a role
of a spring with negative stiffness.}

\section*{Acknowledgements}
The authors are grateful to Prof. D.A.~Indeitsev for useful and stimulating discussions.



\appendix
\section{The Green function, for a beam on the Winkler foundation, in the frequency domain}
\label{app-G-f}
Consider equation
\begin{equation}
 \ddot{G} + G'''' 
 + G = \exp(-\i\Omega t)\delta(x),
\label{beam-Green-alt-maineq}
\end{equation}
where $|\Omega|<1$.
The dispersion relation for the operator in the left-hand side 
of~\eqref{beam-Green-alt-maineq} is 
\begin{equation}
 \omega^4
 -\Omega^2+1=0.
 \label{diss-rel-p=0}
\end{equation}

We look for the solution of this equation in the form of
\begin{gather}
G=2\Re\big(G_0(x,\Omega)\big)
\exp(-\i\Omega t),
\label{beam-spectrum-alt-r-gen}
\\
G_0(x,\Omega)=\left\{
\begin{aligned}	
&\frac C2\,\exp\Big(\i\big(\omega_+ x-\varphi/2\big)\Big),\quad&& x>0; \\
&\frac C2\,\exp\Big(\i\big(\omega_- x+\varphi/2\big)\Big),\quad&& x<0;
\end{aligned}
\right.
\label{beam-spectrum-alt-r}
\end{gather}
where $C,\ \varphi$ are real constants, and $\omega_\pm$ are the wavenumbers 
\begin{gather}
\omega_\pm=\pm a\i+a,
\label{a-a}
\\
a=\frac{\sqrt{2}}{2}
\sqrt[4]{1 - \Omega^2 },
\label{beam-a}
\end{gather} 
which satisfy dispersion relation \eqref{diss-rel-p=0}.
Calculating the real parts of jumps for the first, the second, and the third derivative of
the right-hand side of 
\eqref{beam-spectrum-alt-r-gen} results in the following set of boundary
conditions at $x=0$:
\begin{gather}
\Re[\i\omega]=0,
\label{beam-bc1-s}
\\
\Re[(\i\omega)^2]=0,
\label{beam-bc2-s}
\\
2C\Re[(\i\omega)^3]=1.
\label{beam-bc3-s}
\end{gather}
The left-hand side of Eq.~\eqref{beam-bc2-s} equals
zero identically.
Calculating the left-hand sides 
of \eqref{beam-bc1-s}, \eqref{beam-bc3-s} yields two equations to define
unknown constants $\varphi$ and~$C$:
%
\begin{gather}
\sin\frac\varphi2-\cos\frac\varphi2=0,\\
a^3\left(-\sin\frac\varphi2+3\sin\frac\varphi2+3\cos\frac\varphi2-\cos\frac\varphi2\right)
=\frac1{2C}.
\end{gather}
Resolving these equations yields:
\begin{gather}
C=
\frac{1}{2
(1-\Omega^2)^{3/4}
}\,
,\\
\tan\frac\varphi2=1.
\end{gather}
Hence, 
\begin{gather}
2\Re G_0(x,\Omega) = 
\frac{1}{2}\,
\frac{\mathrm{e}^{-a|x|}\cos{\big( a|x| -\pi/4 \big)}}
{(1-\Omega^2)^{3/4}}
.
\label{Gr_b}
\end{gather}

\section{The Green function, for a beam on the Winkler foundation, in the time
domain (the fundamental solution)}
\label{app-G-t}
Consider equation
\begin{equation}
 \ddot{\GG} + \GG'''' 
 + \GG = \delta(t)\delta(x).
 \label{green-time}
\end{equation}
According to
\eqref{beam-Green-alt-maineq},
\eqref{beam-spectrum-alt-r-gen},
\eqref{green-time}
the Laplace transform $\GG_L(0,p)$ of $\GG(x,t)\big|_{x=0}$ equals
\begin{equation}
\GG_L(0,p)=2\Re G_0(0,-p\i)=\frac1{2^{3/2}(p^2+1)^{3/4}}.
\end{equation}
Applying the inverse Laplace transform yields \cite{Slepian-Non}
\begin{equation}
\GG(0,t)=\frac{\sqrt\pi}{2^{7/4}\Gamma(3/4)}\,t^{1/4}J_{1/4}(t),
\label{fs-td}
\end{equation}
where $\Gamma(\cdot)$ is the Gamma function (the Euler integral of the second kind),
$J_{1/4}(\cdot)$ is the Bessel function of the first kind of order $1/4$.

\section*{Declaration of interest}
None


\begin{thebibliography}{10}
\expandafter\ifx\csname url\endcsname\relax
  \def\url#1{\texttt{#1}}\fi
\expandafter\ifx\csname urlprefix\endcsname\relax\def\urlprefix{URL }\fi
\expandafter\ifx\csname href\endcsname\relax
  \def\href#1#2{#2} \def\path#1{#1}\fi

\bibitem{ursell1951trapping}
F.~Ursell, Trapping modes in the theory of surface waves, Mathematical
  Proceedings of the Cambridge Philosophical Society 47~(2) (1951) 347--358.

\bibitem{kaplunov1986torsional}
J.~Kaplunov, The torsional oscillations of a rod on a deformable foundation
  under the action of a moving inertial load, Izvestiya Akademii Nauk SSSR, MTT
  (Mechanics of solids) 6 (1986) 174--177, (in Russian).

\bibitem{abramyan1992characteristics}
A.~Abramian, V.~Andreyev, D.~Indeitsev, The characteristics of the oscillations
  of dynamical systems with a load-bearing structure of infinite extent,
  Modelirovaniye v mekhanike 6~(2) (1992) 3--12, (in Russian).

\bibitem{kaplunov1995simple}
J.~Kaplunov, S.~Sorokin, A simple example of a trapped mode in an unbounded
  waveguide, The Journal of the Acoustical Society of America 97 (1995)
  3898--3899.

\bibitem{abramyan1998trapping}
A.~Abramyan, D.~Indeitsev, Trapping modes in a membrane with an inhomogeneity,
  Acoustical Physics 44 (1998) 371--376.

\bibitem{Gavrilov-2006-trans}
S.~Gavrilov, The effective mass of a point mass moving along a string on a
  {W}inkler foundation, PMM Journal of Applied Mathematics and Mechanics 70~(4)
  (2006) 582--589.

\bibitem{gavrilov2002etm}
S.~Gavrilov, D.~Indeitsev, {The evolution of a trapped mode of oscillations in
  a ``string on an elastic foundation -- moving inertial inclusion'' system},
  PMM Journal of Applied Mathematics and Mechanics 66~(5) (2002) 825--833.

\bibitem{alekseev2002vibration}
V.~Alekseev, D.~Indeitsev, Y.~Mochalova, Vibration of a flexible plate in
  contact with the free surface of a heavy liquid, Technical Physics 47~(5)
  (2002) 529--534.

\bibitem{indeitsev2004localization}
D.~Indeitsev, E.~Osipova, Localization of nonlinear waves in elastic bodies
  with inclusions, Acoustical Physics 50~(4) (2004) 420--426.

\bibitem{kaplunov2008example}
J.~Kaplunov, E.~Nolde, An example of a quasi-trapped mode in a weakly
  non-linear elastic waveguide, Comptes Rendus M\'{e}canique 336~(7) (2008)
  553--558.

\bibitem{motygin2008trapping}
O.~Motygin, On trapping of surface water waves by cylindrical bodies in a
  channel, Wave Motion 45~(7-8) (2008) 940--951.

\bibitem{nazarov2010sufficient}
S.~Nazarov, Sufficient conditions on the existence of trapped modes in problems
  of the linear theory of surface waves, Journal of Mathematical Sciences
  167~(5) (2010) 713--725.

\bibitem{pagneux2013trapped}
V.~Pagneux, Trapped modes and edge resonances in acoustics and elasticity, in:
  R.~Craster, J.~Kaplunov (Eds.), Dynamic Localization Phenomena in Elasticity,
  Acoustics and Electromagnetism, Springer, 2013, pp. 181--223.

\bibitem{gavrilov2016trapped}
S.~Gavrilov, Y.~Mochalova, E.~Shishkina, Trapped modes of oscillation and
  localized buckling of a tectonic plate as a possible reason of an earthquake,
  in: Proc. Int. Conf. Days on Diffraction (DD), 2016, IEEE, 2016, pp.
  161--165, doi: 10.1109/DD.2016.7756834.

\bibitem{kaplunov2005localized}
J.~Kaplunov, G.~Rogerson, P.~Tovstik, Localized vibration in elastic structures
  with slowly varying thickness, The Quarterly Journal of Mechanics and Applied
  Mathematics 58~(4) (2005) 645--664.

\bibitem{Ind-book-R2E}
D.~Indeitsev, N.~Kuznetsov, O.~Motygin, Y.~Mochalova, Localization of linear
  waves, St. Petersburg University, 2007, (in Russian).

\bibitem{indeitsev2000resonance}
D.~Indeitsev, A.~Sergeev, S.~Litvin, Resonance vibrations of elastic waveguides
  with inertial inclusions, Technical Physics 45~(8) (2000) 963--970.

\bibitem{indeitsev2012motion}
D.~Indeitsev, A.~Abramyan, N.~Bessonov, Y.~Mochalova, B.~Semenov, Motion of the
  exfoliation boundary during localization of wave processes, Doklady Physics
  57~(4) (2012) 179--182.

\bibitem{wang2014vibration}
C.~Wang, Vibration of a membrane strip with a segment of higher density:
  analysis of trapped modes, Meccanica 49~(12) (2014) 2991--2996.

\bibitem{indeitsev2015localization}
D.~Indeitsev, T.~Kuklin, Y.~Mochalova, Localization in a {B}ernoulli-{E}uler
  beam on an inhomogeneous elastic foundation, Vestnik of St.~Petersburg
  University: Mathematics 48~(1) (2015) 41--48.

\bibitem{indeitsev2016evolution}
D.~Indeitsev, S.~Gavrilov, Y.~Mochalova, E.~Shishkina, Evolution of a trapped
  mode of oscillation in a continuous system with a concentrated inclusion of
  variable mass, Doklady Physics 61~(12) (2016) 620--624.

\bibitem{gavrilov2017trapped}
S.~Gavrilov, Y.~Mochalova, E.~Shishkina, Evolution of a trapped mode of
  oscillation in a string on the {W}inkler foundation with point inhomogeneity,
  in: Proc. Int. Conf. Days on Diffraction (DD), 2017, IEEE, 2017, pp.
  128--133, doi: 10.1109/DD.2017.8168010.

\bibitem{luongo2001mode}
A.~Luongo, Mode localization in dynamics and buckling of linear imperfect
  continuous structures, Nonlinear Dynamics 25 (2001) 133--156.

\bibitem{abramyan2011oscillations}
A.~Abramyan, S.~Vakulenko, Oscillations of a beam with a time-varying mass,
  Nonlinear Dynamics 63~(1-2) (2011) 135--147.

\bibitem{abramian2014oscillations}
A.~Abramian, W.~van Horssen, S.~Vakulenko, On oscillations of a beam with a
  small rigidity and a time-varying mass, Nonlinear Dynamics 78~(1) (2014)
  449--459.

\bibitem{abramian2017oscillations}
A.~Abramian, W.~van Horssen, S.~Vakulenko, Oscillations of a string on an
  elastic foundation with space and time-varying rigidity, Nonlinear Dynamics
  88~(1) (2017) 567--580.

\bibitem{mciver2003excitation}
P.~McIver, M.~McIver, J.~Zhang, Excitation of trapped water waves by the forced
  motion of structures, Journal of Fluid Mechanics 494 (2003) 141--162.

\bibitem{arXiv:1805.07382}
S.~Gavrilov, E.~Shishkina, Y.~Mochalova, Non-stationary localized oscillations
  of an infinite string, with time-varying tension, lying on the {W}inkler
  foundation with a point elastic inhomogeneity, arXiv:1805.07382.

\bibitem{fedoruk1977}
M.~Fedoruk, The saddle-point method, Nauka, Moscow, 1977, (in Russian).

\bibitem{nayfeh1993}
A.~Nayfeh, Introduction to Perturbation Techniques, Wiley \& Sons, 1993.

\bibitem{nayfehperturbation}
A.~Nayfeh, Perturbation methods, Weily \& Sons, 1973.

\bibitem{Vladimirov1971}
V.~Vladimirov, Equations of Mathematical Physics, Marcel Dekker, New York,
  1971.

\bibitem{Non-Stationary}
S.~Gavrilov, Non-stationary problems in dynamics of a string on an elastic
  foundation subjected to a moving load, Journal of Sound and Vibration 222~(3)
  (1999) 345--361.

\bibitem{feschenko1967eng}
S.~Feschenko, N.~Shkil, L.~Nikolenko, Asymptotic methods in theory of linear
  differential equations, NY: North-Holland, 1967.

\bibitem{book:811629}
K.~Atkinson, An Introduction to Numerical Analysis, Wiley, 1989.

\bibitem{lurie2007introduction}
K.~Lurie, An introduction to the mathematical theory of dynamic materials,
  Springer, New York, 2007.

\bibitem{rousseau2011elements}
M.~Rousseau, G.~Maugin, M.~Berezovski, Elements of study on dynamic materials,
  Archive of Applied Mechanics 81~(7) (2011) 925--942.

\bibitem{chronopoulos2015enhancement}
D.~Chronopoulos, I.~Antoniadis, M.~Collet, M.~Ichchou, Enhancement of wave
  damping within metamaterials having embedded negative stiffness inclusions,
  Wave Motion 58 (2015) 165--179.

\bibitem{pasternak2014chains}
E.~Pasternak, A.~Dyskin, G.~Sevel, Chains of oscillators with negative
  stiffness elements, Journal of Sound and Vibration 333~(24) (2014)
  6676--6687.

\bibitem{oyelade2017dynamics}
A.~Oyelade, Z.~Wang, G.~Hu, Dynamics of 1d mass--spring system with a negative
  stiffness spring realized by magnets: Theoretical and experimental study,
  Theoretical and Applied Mechanics Letters 7~(1) (2017) 17--21.

\bibitem{grekova2018dd}
E.~Grekova, Harmonic waves in the simplest reduced {K}elvin’s and gyrostatic
  media under an external body follower torque, in: Proc. Int. Conf. Days on
  Diffraction (DD), 2018, IEEE, 2018, pp. 142--148.

\bibitem{huang2014vibration}
X.~Huang, X.~Liu, J.~Sun, Z.~Zhang, H.~Hua, Vibration isolation characteristics
  of a nonlinear isolator using euler buckled beam as negative stiffness
  corrector: A theoretical and experimental study, Journal of Sound and
  Vibration 333~(4) (2014) 1132--1148.

\bibitem{ahn2014active}
L.~Danh, K.~Ahn, Active pneumatic vibration isolation system using negative
  stiffness structures for a vehicle seat, Journal of Sound and Vibration
  333~(5) (2014) 1245--1268.

\bibitem{li2013negative}
Q.~Li, Y.~Zhu, D.~Xu, J.~Hu, W.~Min, L.~Pang, A negative stiffness vibration
  isolator using magnetic spring combined with rubber membrane, Journal of
  Mechanical Science and Technology 27~(3) (2013) 813--824.

\bibitem{wu2014analysis}
W.~Wu, X.~Chen, Y.~Shan, Analysis and experiment of a vibration isolator using
  a novel magnetic spring with negative stiffness, Journal of Sound and
  Vibration 333~(13) (2014) 2958--2970.

\bibitem{Slepian-Non}
L.~Slepyan, Non-Stationary Elastic Waves, Sudostroenie, Leningrad, 1972, (in
  {R}ussian).

\end{thebibliography}

\end{document}